\documentclass[a4paper, 10pt, conference]{ieeeconf}
\usepackage{cite,subfigure,psfrag,graphicx,amssymb}
\usepackage{tikz}
\usetikzlibrary{calc}
\usepackage[tbtags]{amsmath}
\usepackage{epsfig}
\usepackage{stfloats}
\usepackage{array}
\usepackage{color}
\usepackage{pgfplots}
\pgfplotsset{compat=newest}
\pgfplotsset{plot coordinates/math parser=false}
\newcommand{\E}{\mathrm{E}}
\newcommand{\Prob}{\mathrm{Pr}}

\newcommand{\av}{{\bf a}}

\newcommand{\rv}{{\bf r}}

\newcommand{\Am}{{\bf A}}

\newcommand{\Cm}{{\bf C}}

\begin{document}

\title{On the Effect of Correlated Measurements on the Performance of Distributed Estimation}
\author{\large Mohammed~F.~A.~Ahmed$^{\dagger}$, Tareq~Y.~Al-Naffouri$^{\dagger,\ddagger}$, and Mohamed-Slim Alouini$^{\dagger}$ 
\\ $^\dagger$ King Abdullah University of Science and Technology (KAUST), Thuwal, Saudi Arabia.
\\ $^\ddagger$ King Fahd University of Petroleum and Minerals, Dhahran, Saudi Arabia.\\ \normalsize Email: \{m.ahmed, tareq.alnaffouri, slim.alouini\}@kaust.edu.sa}

\maketitle

\begin{abstract}
We address the distributed estimation of an unknown scalar parameter in Wireless Sensor Networks (WSNs). Sensor nodes transmit their noisy observations over multiple access channel to a Fusion Center (FC) that reconstructs the source parameter. The received signal is corrupted by noise and channel fading, so that the FC objective is to minimize the Mean-Square Error (MSE) of the estimate. In this paper, we assume sensor node observations to be correlated with the source signal and correlated with each other as well. The correlation coefficient between two observations is exponentially decaying with the distance separation. The effect of the distance-based correlation on the estimation quality is demonstrated and compared with the case of unity correlated observations. Moreover, a closed-form expression for the outage probability is derived and its dependency on the correlation coefficients is investigated. Numerical simulations are provided to verify our analytic results.
\end{abstract}

\section{Introduction}
 
Recent advances of micro-sensor fabrication technology allow for producing cheap and small sensor nodes with wireless communication capabilities. Consequently, Wireless Sensor Networks (WSNs) become an economically sound solution to wide range of applications such as environmental and wildlife habitat monitoring, target tracking for defense purposes, and health care \cite{Akyildiz2002}. Typical WSN consists of large number of sensor nodes deployed in an area of interest to collect specific information about the surrounding environment. The need for large number of sensor nodes in WSNs while being cost-effective constraints the industry standards to produce battery-powered sensor nodes with simple hardware. As a result of the limited energy and processing capabilities of sensor nodes, the collected information has to be sent to a Fusion Center (FC) for centralized processing. 

One important application of WSNs is the distributed estimation of scalar parameters (see, e.g., \cite{J_Giannakis2006}, and references therein). In such application, sensor nodes transmit their observations over a Multiple Access Channel (MAC) to the FC. The received signal is distorted by the channel fading and the additive noise. The FC is required to reconstruct the source parameter with minimum Mean-Square Error (MSE). 
Depending on the available information about the source statistics, different estimators can be used to achieve the MSE criterion. The performance of the Best Linear Unbiased Estimation (BLUE) \cite{J_Goldsmith2007}, Minimum Mean Squared Error (MMSE) estimator \cite{J_Cihan2008,J_Dey2011c}, and Maximum Likelihood Estimator (MLE) \cite{J_Aysal2008,J_Wang2011} are studied in literature. Both orthogonal MAC \cite{J_Bahceci2008} and coherent MAC \cite{J_Goldsmith2008,J_Chaudhary2012} are considered in the distributed estimation problem. 
Assuming Gaussian source signal and noise, amplify-and-forward schemes significantly outperform the traditional source-channel coding for both multiple access channels \cite{J_Gastpar2003a}.
Optimal power allocation for sensor nodes under different constraints is addressed in \cite{J_Goldsmith2007,J_Cihan2008,J_Bahceci2008}. 
Asymptotic behavior of the distortion is also studied in \cite{J_Dey2011b}. 
The MSE performance for the coherent MAC asymptotically approaches to zero as the number of sensors increases to infinity. However, this is not the case for the orthogonal MAC where the MSE reaches a finite non-zero value as the number of sensor nodes increases \cite{J_Goldsmith2008}. Diversity order of estimation distortion is introduced in \cite{J_Goldsmith2007} and shown to be given by the number of sensors.

In most WSN applications, the source parameter is a physical quantity like temperature, pressure, humidity, sound, ... etc. Therefore, the sensor node observations are correlated where the correlation coefficient is exponentially decaying with distance. 
In literature, simple signal models were usually assumed. For example, unity correlated observations are assumed in \cite{J_Goldsmith2007}, 
where sensor nodes measure a noisy version of the source signal. The correlation between the observations and the source signal (denoted hereafter as the source-node correlation) is considered in \cite{J_Goldsmith2007}. However, in this model the correlation between observations (denoted hereafter as the inter-node correlation) is determined by the source-node correlation as we will show later. Considering a correlation model with inter-node correlation that determined by the distance between sensor nodes is a more realistic assumption \cite{Conf_AlMurad2010,J_Chaudhary2012}. 

In this paper, we study the distributed estimation of a scalar parameter where sensor nodes transmit their observations to the FC over a coherent MAC. The observations are spatially correlated and corrupted by noise. Moreover, the communication channel is subject to fading and Additive White Gaussian Noise (AWGN). The FC uses the received signal to estimate the source parameter using LMMSE estimator. The distance-based correlation model of \cite{J_Akyildiz2006} is used in this paper to characterize the source-node correlation and the inter-node correlation. 
The effect of the distance-based correlation on the estimation performance is demonstrated and compared with the case of unity correlated observations. The outage probability is adopted as the performance measure. A new closed-form expression for the outage probability in terms of quadratic forms is introduced. It is shown that less correlated observations degrade the performance. 

Hereafter, small letters, bold small letters, and bold capital letters will
designate scalars, vectors, and matrices, respectively. If $\Am$ is a
matrix, then $\Am^H$, $\Am^T$, and $\text{eig} \left( \Am \right)$ denote the hermitian, the transpose, and the eigenvalues of $\Am$, respectively. We define $\text{diag} ( \av )$ to be a diagonal matrix formed from vector $\av$.
The function $\lfloor \cdot \rfloor$ returns the same value for the positive values and zero for the negative values and the function $\text{max} ( \cdot )$ returns the maximum value.

\begin{table*}[b!]
\setcounter{equation}{5}
\hrule 
\vspace{2.4 mm}
\begin{equation}
\tilde{D} =   \frac{   \sigma_s^2 \left( \sum\limits_{i=1}^{N} \sum\limits_{j=1}^{N} a_i a_j g_i g_j \rho_{ij} - \left( \sum\limits_{i=1}^{N} a_i  g_i \rho_i \right)^2  \right)+ \sigma_n^2 \sum\limits_{i=1}^{N} a_i^2  g_i^2  + \sigma_\nu^2 }{  \sigma_s^2 \sum\limits_{i=1}^{N} \sum\limits_{j=1}^{N} a_i a_j g_i g_j \rho_{ij} + \sigma_n^2 \sum\limits_{i=1}^{N} a_i^2  g_i^2  + \sigma_\nu^2}. 
\label{NormDes}
\end{equation}
\end{table*}
\setcounter{equation}{0}

\section{System Model}

Consider a WSN consisting of $N$ sensor nodes and a FC as shown in Fig.~1. Sensor nodes are required to observe a scalar parameter modeled by a zero-mean complex Gaussian random variable $s \sim {\cal CN} (0, \sigma^2_s)$. The signals measured by individual sensor nodes can be described as
\begin{eqnarray}
x_i = s_i + n_i, \quad i = 1,\dots,N,
\label{Eq:ObservatioModel}
\end{eqnarray}
\noindent where $s_i$ is the $i$th sensor node observation and $n_i \sim {\cal CN} (0, \sigma_n^2)$ is the observation noise. The signals $s$ and $s_i, i = 1,\dots, N$, are modeled as zero-mean joint Gaussian random variables, i.e. $\E \{ s_i \} = 0, i = 1,\dots,N$, $\E \{ s s_i\} = \rho_{i} \sigma^2_s, i = 1,\dots, N$, where $\rho_i$ is the source-node correlation coefficient between $s$ and $s_i$. Moreover, sensor node observations $s_i, i = 1,\dots, N$, are correlated to each other, i.e. $\E \{ s_i s_j\} = \rho_{ij} \sigma^2_s, i,j = 1,\dots, N, i \neq j$, where $\rho_{ij}$ is the inter-node correlation coefficient between $s_i$ and $s_j$. 

The correlation coefficients are non-negative and decrease monotonically with distance. 
Using the power exponential model presented in \cite{J_Akyildiz2006}, $\rho_i$ and $\rho_{ij}$ are functions of $d_i$ and $d_{ij}$, respectively, according to the relation 
\begin{eqnarray}
\rho (d) = e^{-{\left(\frac{d}{\theta_1}\right)}^{\theta_2}},  \theta_1 > 0,   0 < \theta_2 \le 2, d \in \{ d_{i}, d_{ij} \},
\end{eqnarray}
\noindent where $d_i$ is the distance between the event source and the $i$th node, $d_{ij}$ is distance between sensor $i$ and $j$, $\rho_{ii} = 1$, and $\theta_1$ and $\theta_2$ are the model parameters, where $\theta_1$ normalizes the distance and $\theta_2$ controls the correlation decay rate. Let us define the matrix $\Cm$ as the cross-node correlation matrix with $\rho_{ij}$ is the element at the $i$th row and $j$th column. Assuming random sensor node locations, $\Cm$ will be a full-rank matrix, and therefore this model will be referred to as {\it full-rank} correlation model. The aforementioned model for the correlation coefficients $\rho_{i}$ and $\rho_{ij}$ is more generalized than the one used in \cite{J_Goldsmith2007,J_Cihan2008,J_Dey2011c} where the signal is given as $
x_i {(t)} = s {(t)} + v_i {(t)} , i = 1,\dots,N$, and thus $s_i {(t)} = s {(t)} $ in this case. This results in {\it unity} correlated source and observations, i.e. $\rho_{i} = 1$ and $\rho_{ij} = 1$, $i,j = 1,\dots, N$. Also, the signal model used in \cite{J_Goldsmith2008,J_Dey2011a} is $x_i {(t)} = h_i s{(t)} + v_i {(t)} , i = 1,\dots,N,$ which corresponds to $s_i {(t)} = h_i s {(t)}$  in Eq.~(\ref{Eq:ObservatioModel}) and results in the special case  $\rho_{ij} = \rho_{i}\rho_{j}$ where $ \E \left\{ s {(t)} s_i {(t)} \right\} = \sigma_s^2 h_i = \sigma_s^2 \rho_{i}$ and $ \E \left\{ s_i {(t)} s_j {(t)} \right\} = \sigma_s^2 h_i h_j = \sigma_s^2 \rho_{i} \rho_{j}$. Let us define the vector $\rv = \left[ \rho_1 \ \rho_2 \ \dots  \ \rho_N \right]^{T}$, then the cross-node correlation matrix for the later case is given by $\Cm = \rv \rv^T$ which is a rank one matrix and thus will be referred to as {\it rank-one} correlation model. In the rest of the paper, we will compare between the three correlation models.

For the amplify-and-forward scheme, the transmitted signal from the $i$th sensor node is given by $y_i  = a_i  x_i, i = 1, \dots, N$, where $a_i$ is the amplification factor. Here, the transmit power for each node is $\E \{ |y_i|^2 \} = a_i^2 ( \sigma^2_s + \sigma^2_n )$. Assume that the Channel State Information (CSI) is available at the FC. Then, the coherent combining of the transmitted signals received at the FC is
\begin{eqnarray}
z  = \sum\limits_{i=1}^{N} a_i  g_i x_i + \nu,
\end{eqnarray}
\noindent where $\nu$ is the communication noise $ \nu \sim {\cal CN} (0, \sigma^2_\nu)$ and $g_i$ is the Rayleigh fading for the $i$th node $g_i \sim {\cal CN} (0, \sigma^2_g)$. 

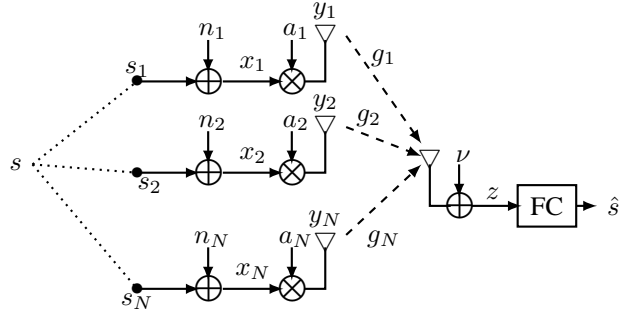
\begin{figure}[t!]
\hspace{0pt}
\begin{tikzpicture}[scale = 0.55]
\tikzstyle{every node}=[font=\normalsize]

\coordinate (s) at (2.5,0);
\coordinate (s1) at (5,2);
\coordinate (s2) at (5,-0.2);
\coordinate (sN) at (5,-3);
\coordinate (r) at (12,0);
\coordinate (s1_shift) at (1.5,0);
\coordinate (s2_shift) at (1.5,0);
\coordinate (sN_shift) at (1.5,0);

\draw ($(s)+(-0.4,0)$) node {$s$};
\draw ($(s1)+(0,0.3)$) node {$s_1$};
\draw ($(s2)+(0.3,-0.3)$) node {$s_2$};
\draw ($(sN)+(0,-0.3)$) node {$s_N$};


\draw[black, dotted,   thick, rotate=0] (s)  -- (s1) node [midway,above] {}(s1) node {$\bullet$};
\draw[black, dotted,   thick, rotate=0] (s)  -- (s2) node [pos=0.4,above] {}(s2) node {$\bullet$};
\draw[black, dotted,   thick, rotate=0] (s)  -- (sN) node [midway,above] {}(sN) node {$\bullet$};


\draw [black,  thick,-latex] (s1) -- ($(s1)+(s1_shift)$) node [right = -6] {$\bigoplus$}; 
\draw ($(s1)+(s1_shift)+(0.3,1.2)$) node {$n_1$}; 
\draw [black,  thick,-latex] ($(s1)+(s1_shift)+(0.2,1)$) -- ($(s1)+(s1_shift)+(0.2,0.2)$); 
\draw [black,  thick,-latex] ($(s1)+(s1_shift)+(0.55,0)$) -- ($(s1)+(s1_shift)+(2,0)$)  node [midway,above] {$x_1$} node [right = -6] {$\bigotimes$}; 
\draw ($(s1)+(s1_shift)+(3.3,1.2)+(-1,0)$) node {$a_1$}; 
\draw [black,  thick,-latex] ($(s1)+(s1_shift)+(3.2,1)+(-1,0)$) -- ($(s1)+(s1_shift)+(3.2,0.2)+(-1,0)$); 
\draw [black,  thick] ($(s1)+(s1_shift)+(3.5,0)+(-1,0)$) -- ($(s1)+(s1_shift)+(6,0)+(-1,0)+(-2,0)$) -- ($(s1)+(s1_shift)+(6,1)+(-1,0)+(-2,0)$) node [above = -5] {$\bigtriangledown$}  node [above = 4] {$y_1$}; 
\draw [black, dashed, thick,-latex] ($(s1)+(s1_shift)+(6.5,1.1)+(-1,0)+(-2,0)$) -- ($(r)+(-0.2,0.5)$)  node [pos=0.2, right] {$g_1$}; 

\draw [black,  thick,-latex] (s2) -- ($(s2)+(s2_shift)$) node [right = -6] {$\bigoplus$}; 
\draw ($(s2)+(s2_shift)+(0.3,1.2)$) node {$n_2$}; 
\draw [black,  thick,-latex] ($(s2)+(s2_shift)+(0.2,1)$) -- ($(s2)+(s2_shift)+(0.2,0.2)$); 
\draw [black,  thick,-latex] ($(s2)+(s2_shift)+(0.55,0)$) -- ($(s2)+(s2_shift)+(2,0)$)   node [midway,above] {$x_2$} node [right = -6] {$\bigotimes$}; 
\draw ($(s2)+(s2_shift)+(3.3,1.2)+(-1,0)$) node {$a_2$}; 
\draw [black,  thick,-latex] ($(s2)+(s2_shift)+(3.2,1)+(-1,0)$) -- ($(s2)+(s2_shift)+(3.2,0.2)+(-1,0)$); 
\draw [black,  thick] ($(s2)+(s2_shift)+(3.5,0)+(-1,0)$) -- ($(s2)+(s2_shift)+(6,0)+(-1,0)+(-2,0)$) -- ($(s2)+(s2_shift)+(6,1)+(-1,0)+(-2,0)$) node [above = -5] {$\bigtriangledown$}  node [above = 4] {$y_2$}; 
\draw [black, dashed, thick,-latex] ($(s2)+(s2_shift)+(6.5,1.1)+(-1,0)+(-2,0)$) -- ($(r)+(-0.2,0.2)$)  node [pos=0.3, above] {$g_2$}; 

\draw [black,  thick,-latex] (sN) -- ($(sN)+(sN_shift)$) node [right = -6] {$\bigoplus$}; 
\draw ($(sN)+(sN_shift)+(0.3,1.2)$) node {$n_N$}; 
\draw [black,  thick,-latex] ($(sN)+(sN_shift)+(0.2,1)$) -- ($(sN)+(sN_shift)+(0.2,0.2)$); 
\draw [black,  thick,-latex] ($(sN)+(sN_shift)+(0.55,0)$) -- ($(sN)+(sN_shift)+(2,0)$)   node [midway,above] {$x_N$}  node [right = -6] {$\bigotimes$}; 
\draw ($(sN)+(sN_shift)+(3.3,1.2)+(-1,0)$) node {$a_N$}; 
\draw [black,  thick,-latex] ($(sN)+(sN_shift)+(3.2,1)+(-1,0)$) -- ($(sN)+(sN_shift)+(3.2,0.2)+(-1,0)$); 
\draw [black,  thick] ($(sN)+(sN_shift)+(3.5,0)+(-1,0)$) -- ($(sN)+(sN_shift)+(6,0)+(-1,0)+(-2,0)$) -- ($(sN)+(sN_shift)+(6,1)+(-1,0)+(-2,0)$) node [above = -5] {$\bigtriangledown$}  node [above = 4] {$y_N$}; 
\draw [black, dashed, thick,-latex] ($(sN)+(sN_shift)+(6.5,1.3)+(-1,0)+(-2,0)$) -- ($(r)+(-0.2,0)$)  node [pos=0.5,below = 8] {$g_N$};


\draw ($(r)+(0,0.2)$) node {$\bigtriangledown$};
\draw ($(r)+(1.5,-0.7)$) node {$z$};
\draw [black,   thick] (r) -- ($(r)+(0,-1)$) -- ($(r)+(0.5,-1)$) node [right = -6] {$\bigoplus$};  
\draw ($(r)+(0.8,0.2)$) node {$\nu$}; 
\draw [black,   thick,-latex]  ($(r)+(0.7,0)$) --  ($(r)+(0.7,-0.8)$);
\draw [black,   thick,-latex] ($(r)+(1,-1)$) -- ($(r)+(2.2,-1)$);
\draw [black,   thick] ($(r)+(2.1,-.5)$) rectangle ($(r)+(3.5,-1.5)$) ($(r)+(2.8,-1)$) node {FC};
\draw [black,   thick,-latex] ($(r)+(3.5,-1)$) -- ($(r)+(4,-1)$) [right] node {$\hat{s}$}; 
\end{tikzpicture}
\caption{System model.}  
\end{figure}

Given the signal and channel statistics, the LMMSE estimate $\hat{s}$ can be expressed as
\begin{eqnarray}
\!\!\!\! \hat{s} \!  = \!  \frac{\E \left\{ z s \right\}}{\E \left\{ z^2\right\}} z \! 
= \!  \frac{  z \sigma_s^2 \sum\limits_{i=1}^{N} a_i  g_i \rho_i }{  \sigma_s^2 \sum\limits_{i=1}^{N} \sum\limits_{j = 1}^{N}  a_i a_j     g_i g_j \rho_{ij} +  \sum\limits_{i=1}^{N} \sigma_n^2 a_i^2  g_i^2  + \sigma_\nu^2} 
 \ 
\end{eqnarray}
\noindent and the corresponding distortion becomes
\begin{eqnarray}
 D  &=&  \E \left\{ ( s - \hat{s} )^2  \right\} = \sigma_s^2 - \frac{ \left( \E \left\{ z s \right\} \right)^2}{\E \left\{z^2\right\}}  = \sigma_s^2   \nonumber \\
  &-&  \frac{ \sigma_s^4 \left( \sum\limits_{i=1}^{N} a_i  g_i \rho_i \right)^2 }{  \sigma_s^2 \sum\limits_{i=1}^{N} \sum\limits_{j=1}^{N} a_i a_j      g_i g_j \rho_{ij} + \sigma_n^2 \sum\limits_{i=1}^{N} a_i^2  g_i^2  + \sigma_\nu^2}.
\end{eqnarray}

The normalized distortion $\tilde{D} = \frac{D}{\sigma_s^2}$ is then given by Eq.~(\ref{NormDes}) (shown at the bottom of the next page). This expression is the generalization of the equivalent one given in \cite[Eq.~2]{J_Dey2011c} for the unity correlated source and observations. The first term in the numerator and the first term of the denominator in Eq.~(\ref{NormDes}) are the result of the inter-node correlation.

\section{Effect of the correlation on the distortion}

In this section, the special cases of unity correlation and rank-one model are compared to full-rank model. For simplicity, the channel fading is neglected, i.e. $g_i = 1, i = 1,\dots, N$, and Equal Power Allocation (EPA) is assumed, i.e. $a_i = \sqrt{ P_{\rm tot} / N (\sigma^2_s + \sigma^2_n) } = a$, where $P_{\rm tot}$ is the total transmit power for all sensor nodes. Accordingly, the normalized distortion expression for the full-rank model reduces to 
\setcounter{equation}{7}
\begin{eqnarray}
\tilde{D}_{0}^{\rm FR} =  \hspace{5.5cm} \nonumber \\  \frac{  \sigma_s^2 a^2  \left(  \sum\limits_{i=1}^{N} \sum\limits_{j=1}^{N} \rho_{ij} - \left( \sum\limits_{i=1}^{N} \rho_{i} \right)^2 \right) + N \sigma_n^2 a^2 + \sigma_\nu^2 }{  \sigma_s^2 a^2
  \left(  \sum\limits_{i=1}^{N} \sum\limits_{j=1}^{N} \rho_{ij} \right) + N \sigma_n^2 a^2  + \sigma_\nu^2}. 
\label{AsymptoticD1}    
\end{eqnarray}

Considering the signal model in \cite[Eq.~10]{J_Goldsmith2008}, which results in the special case  $\rho_{ij} =  \rho_{i} \rho_{j}$ (rank-one model), the corresponding normalized distortion is given by
\begin{eqnarray}
\tilde{D}_{0}^{\rm RO} &=&  \frac{    \frac{P_{\rm tot} \sigma_n^2}{\left(  \sigma_s^2 + \sigma_n^2 \right)}   + \sigma_\nu^2 }{  \frac{P_{\rm tot} \sigma_s^2}{N \left(  \sigma_s^2 + \sigma_n^2 \right)}  \left(  \sum\limits_{i=1}^{N}\rho_{i} \right)^2 + \frac{P_{\rm tot} \sigma_n^2}{\left(  \sigma_s^2 + \sigma_n^2 \right)}  + \sigma_\nu^2}. 
\label{AsymptoticD2}    
\end{eqnarray}

Finally, for unity correlated source and observations, i.e. $\rho_{i} = 1$ and $\rho_{ij} = 1, \ \forall i,j$, the normalized distortion expression reduces to
\begin{eqnarray}
\tilde{D}_{0}^{\rm U} &=&  \frac{    \frac{P_{\rm tot} \sigma_n^2}{ \left(  \sigma_s^2 + \sigma_n^2 \right)} + \sigma_\nu^2 }{   \frac{ P_{\rm tot} }{ \left(  \sigma_s^2 + \sigma_n^2 \right)}  (N \sigma_s^2 + \sigma_n^2)  + \sigma_\nu^2}. 
\label{AsymptoticD3}    
\end{eqnarray}
\noindent which is equivalent to the expression in \cite[Eq.~2]{J_Dey2011c}. 

Comparing the aforementioned expressions, it is apparent that $\tilde{D}_{0}^{\rm RO}$ and $\tilde{D}_{0}^{\rm U} \rightarrow 0$ as the number of sensor nodes goes to infinity. However, $\tilde{D}_{0}^{\rm FR}$ does not vanish under the same condition. The effect of correlation on the distortion when the number of sensor nodes increases is depicted in Fig.~2. Eqs. (\ref{AsymptoticD1}), (\ref{AsymptoticD2}), and (\ref{AsymptoticD3}) are averaged over 1000 random realization of sensor node locations and plotted for increasing $N$. Here, {the observation Signal-to-Noise Ratio (SNR) is defined as $\sigma^2_s / \sigma^2_n$ = 20~dB and the communication SNR as $(\sigma^2_s +\sigma^2_n) / \sigma^2_\nu$  = 20~dB, where $\sigma_s^2  = 1$. The correlation model has $\theta_1$ = 250 and $\theta_2$ = 1. The total transmit power of all sensor nodes is $P_{\rm tot}$ = 10 dB (normalized to $\sigma^2_s$).} As expected, the distortion for the unity correlated case tends to zero as $N$ increases. The same behavior is noticed for the rank-one model, however with slightly higher distortion in this case. Conversely, the distortion for full-rank model exhibit a floor behavior at a non-zero distortion ($\approx$ 0.182). Therefore, correlation between observations should be taken into consideration when designing distributed estimation schemes because weak correlation degrades its performance.

\section{Outage probability analysis}

The normalized distortion in Eq.~(\ref{NormDes}) depends on the channel fading, correlation coefficients, channel and measurement noise power, and amplification factors. Given the distance-based correlation model, the correlation coefficients will be invariable for specific network geometry. Therefore, with fixed power allocation and noise power, the channel fading is the only determining factor for the fluctuation in the estimation distortion. In many applications, the interest is to characterize the maximum distortion rather than the average distortion. In such situations, the outage probability is used as a performance measure where the outage refers to the event at which a desired performance level cannot be satisfied. Here, the outage probability is defined as the probability that the normalized distortion exceeds a certain value $ \tilde{\delta}$,
\begin{eqnarray}
P_{\rm out} (\tilde{\delta}) = \Prob  (\tilde{D}  \ge \tilde{\delta}) .
\label{PoutDef} 
\end{eqnarray}

\begin{figure}[t!]
\centering
\newlength\figureheight 
\newlength\figurewidth 
\setlength\figureheight{0.309\textwidth}
\setlength\figurewidth{0.4\textwidth}
\begin{tikzpicture}

\begin{semilogxaxis}[%
view={0}{90},
width=\figurewidth,
height=\figureheight,
scale only axis,
y tick label style={/pgf/number format/.cd,%
 /pgf/number format/precision=4,
 scaled x ticks = false,
          fixed},
xmin=1, xmax=1000,
xminorticks=true,
xlabel={$N$},
xmajorgrids,
xminorgrids,
ymin=0, ymax=0.03,
ylabel={$\E \{\tilde{D}_{0} \}$},
ymajorgrids]
\addplot [
color=blue,
dash pattern=on 1pt off 3pt on 3pt off 3pt,
line width=1.0pt,
mark=o,
mark options={solid}
]
coordinates{
 (1,0.0251791482589066)(2,0.00956599672989086)(3,0.00567839542241515)(4,0.00399058656421733)(6,0.00248359865890106)(8,0.00179503128200769)(10,0.00140503818355683)(20,0.000670618862812041)(40,0.000327184539651787)(100,0.000128925874789742)(200,6.41445400510856e-005)(300,4.26873245548269e-005)(400,3.19912768493655e-005)(500,2.55782093992717e-005)(600,2.1307856737177e-005)(700,1.8260812651337e-005)(800,1.59750993671284e-005)(900,1.41971656547517e-005)(1000,1.27769839861703e-005) 
};\label{plot:One}

\addplot [
color=black,
solid,
line width=1.0pt,
mark=+,
mark options={solid}
]
coordinates{
 (1,0.0198009999990202)(2,0.00749367091120257)(3,0.00444691439935902)(4,0.00312774899205307)(6,0.00194623257020382)(8,0.00140740703871407)(10,0.00110079691079634)(20,0.000525226492165301)(40,0.000256309913385765)(100,0.00010100989597021)(200,5.02524995593789e-005)(300,3.3445559134939e-005)(400,2.50631280738665e-005)(500,2.00404023742234e-005)(600,1.66947240593135e-005)(700,1.4306327979113e-005)(800,1.25157824157301e-005)(900,1.11235812028364e-005)(1000,1.00101007968791e-005) 
};\label{plot:Two}

\end{semilogxaxis}

\begin{semilogxaxis}[%
view={0}{90},
width=0.7\figurewidth,
height=0.7\figureheight,
xshift=0.27\figurewidth,
yshift=0.27\figureheight,
y tick label style={/pgf/number format/.cd,%
 scaled x ticks = false,
          fixed},
scale only axis,
axis background/.style={fill=white},
scale only axis,
xmin=1, xmax=1000,
xminorticks=true,
xmajorgrids,
xminorgrids,
ymin=0.18, ymax=0.24,
ymajorgrids,
legend style={draw=black,fill=white,align=left}]
\addplot [
color=red,
dotted,
line width=1.0pt,
mark=diamond,
mark options={solid}
]
coordinates{
 (1,0.231817409991678)(2,0.207498365199716)(3,0.199140123684108)(4,0.194482070988158)(6,0.190717846584556)(8,0.188465809994503)(10,0.187628460140093)(20,0.185798436574692)(40,0.1845715668613)(100,0.18378372086709)(200,0.183594691704024)(300,0.183425638477268)(400,0.183446234200772)(500,0.183381883126919)(600,0.183373719673596)(700,0.183400274017953)(800,0.183388671376955)(900,0.183330399227119)(1000,0.183376403088507) 
}; \label{plot:Three}
\addlegendentry{Full-rank model};
\addlegendimage{/pgfplots/refstyle=plot:One}\addlegendentry{Rank-one model}
\addlegendimage{/pgfplots/refstyle=plot:Two}\addlegendentry{Unity correlation}

\end{semilogxaxis}
\end{tikzpicture}%
\caption{The behavior of the normalized distortion as the number of sensor nodes $N$ increases for the three correlation models
($\theta_1$ = 250, $\theta_2$ = 1, $\sigma^2_s / \sigma^2_n$ = 20~dB, $\sigma^2_x / \sigma^2_\nu$  = 20~dB, $\sigma_s^2  = 1$, and $P_{\rm tot}$ = 10 dB).}
\end{figure}
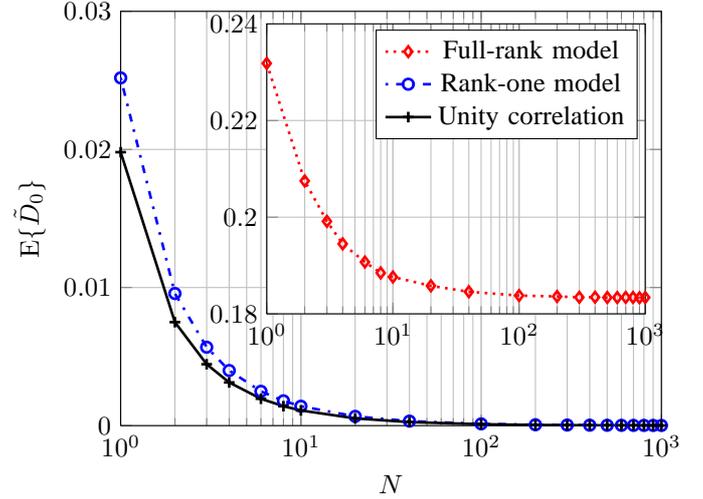

To find a closed-form expression for the outage probability, we first express the distortion in vector form. Let us define the vectors $\mathbf{g} = \left[ g_1 \ g_2 \ \dots \ g_N \right]^{T}$, $\mathbf{a} = \left[ a_1 \ a_2 \ \dots \ a_N \right]^{T}$, and $\mathbf{r}$ and $\mathbf{C}$ as defined previously. It follows that $\sum\limits_{i=1}^{N} a_i  g_i \rho_i = \mathbf{z}^{H} \mathbf{g}$, where $\mathbf{z} =   \mathbf{W}\mathbf{r}$ and $\mathbf{W} =   \text{diag} \left(\mathbf{a} \right)$. Similarly, $\left( \sum\limits_{i=1}^{N} a_i  g_i \rho_i \right)^2 = \left( \mathbf{z}^{H} \mathbf{g} \right)^2 = \mathbf{g}^{H} \mathbf{F} \mathbf{g} = ||\mathbf{g}||^2_{\mathbf{F}}$, where $\mathbf{F} = \mathbf{z} \mathbf{z}^{H}$. Also, $\sum\limits_{i=1}^{N} \sum\limits_{j=1}^{N} a_i a_j     g_i g_j \rho_{ij} =  \mathbf{g}^{H} \mathbf{B} \mathbf{g} =  ||\mathbf{g}||^2_{\mathbf{B}}$, where $\mathbf{B} = \mathbf{W} \mathbf{C} \mathbf{W}^{T}$ and $
\sum\limits_{i=1}^{N} a_i^2  g_i^2  = \mathbf{g}^{H} \mathbf{W}^2 \mathbf{g} = ||\mathbf{g}||^2_{\mathbf{W}^2}$. Then the normalized distortion can be expressed in terms of indefinite quadratic forms as
\begin{eqnarray}
\tilde{D}  =   \frac{ ||\mathbf{g}||^2_{\mathbf{B}_1} +  \sigma^2_\nu}{ ||\mathbf{g}||^2_{\mathbf{B}_2}  +  \sigma^2_\nu },
\label{IndQforms} 
\end{eqnarray}
\noindent where $\mathbf{B}_1 = \sigma^2_s \left( \mathbf{B} - \mathbf{F} \right) + \sigma^2_n \mathbf{W}^2$ and $ \mathbf{B}_2  = \sigma^2_s \mathbf{B} + \sigma_n^2 \mathbf{W}^2$. Accordingly, the outage probability takes the form
\begin{eqnarray}
P_{\rm out} (\tilde{\delta})  &=& \Prob \left(  \frac{ ||\mathbf{g}||^2_{\mathbf{B}_1} +  \sigma^2_\nu}{ ||\mathbf{g}||^2_{\mathbf{B}_2}  +  \sigma^2_\nu } \ge \tilde{\delta} \right) \nonumber\\
&=& \Prob \left( \tilde{\delta} \ ||\mathbf{g}||^2_{\mathbf{B}_2} - ||\mathbf{g}||^2_{\mathbf{B}_1}      \le    \sigma^2_\nu   - \sigma^2_\nu  \tilde{\delta}\right)  \ \nonumber \\
&=& \Prob \left(  ||\mathbf{g}||_{\mathbf{E}(\tilde{\delta})} \le  \left( 1 - \tilde{\delta}  \right) \sigma^2_\nu  \right) ,
\end{eqnarray}
\noindent where $\mathbf{E}(\tilde{\delta}) =  \tilde{\delta} \ \mathbf{B}_2 - \mathbf{B}_1 $. Using the results of \cite{AlNaffouri2009}, the outage probability $P_{\rm out} ( \tilde{\delta} ) $ can be expressed as
\begin{eqnarray}
P_{\rm out} ( \tilde{\delta}) &=& u \left( \left( 1 - \tilde{\delta}  \right) \sigma^2_\nu \right) + \sum\limits_{l=1}^{N}  \frac{ (-\lambda_l)^N}{ \prod\limits_{i, l \ne i}  ( \lambda_i - \lambda_l )} \frac{1}{\lambda_l} \nonumber\\
&\times & e^{-\frac{\left( 1 - \tilde{\delta}  \right) \sigma^2_\nu}{\lambda_l}} u \left(\frac{\left( 1 - \tilde{\delta}  \right) \sigma^2_\nu}{\lambda_l} \right),
\label{outagePrEqn1} 
\end{eqnarray}
\noindent where $\lambda_l = \lambda_l(\tilde{\delta}), i = 1,\dots,N$, are the eigenvalues of the matrix $\mathbf{E}(\tilde{\delta})$ and $u(\cdot)$ is the Heaviside unit-step function. This expression can be simplified if  $\mathbf{E}$ is substituted by its component matrices,
\begin{eqnarray}
\lambda_l (\tilde{\delta}) &=& \text{eig} \left( \mathbf{E} (\tilde{\delta}) \right) = \text{eig} \left( \tilde{\delta} \ \mathbf{B}_2 - \mathbf{B}_1 \right) \nonumber \\
&=& \text{eig} \left( \sigma^2_s \mathbf{F} - \sigma^2_s ( 1 - \tilde{\delta}  ) \mathbf{B} - \sigma_n^2 ( 1 - \tilde{\delta}  ) \mathbf{W}^2  \right) \nonumber \\
&\approx& \sigma^2_s \text{eig} \left( \mathbf{F} -  ( 1 - \tilde{\delta}  ) \mathbf{B}  \right), \quad l = 1,\dots, N,
\end{eqnarray}
where the last approximation results from considering that $\sigma_n^2 \ll \sigma^2_s $. Recall that the matrix $\mathbf{F}$ is the outer product of the vector $\mathbf{z}$ by itself, hence it is of rank one and positive semidefinite. Moreover, $\mathbf{B}$ is also positive semidefinite matrix. Therefore, $\mathbf{E}$ has only one non-negative eigenvalue and all other eigenvalues can simply canceled off in (\ref{outagePrEqn1}) because of the unit-step function. Finally, the expression of outage probability simplifies to
\begin{eqnarray}
P_{\rm out} (\tilde{\delta}) = 1 - \frac{ \lambda_{+}^{N-1}  e^{-\frac{\left( 1 - \tilde{\delta}  \right) \sigma^2_\nu}{\lambda_{+} }}}{ \prod\limits_{i, l \ne i}  ( \lambda_{+} - \lambda_i)}   , \ 0 \le \tilde{\delta} \le 1,
\label{outagePrEqn} 
\end{eqnarray}
\noindent where $\lambda_{+} = \lambda_{+}(\tilde{\delta})$ is the only non-negative eigenvalue of $\mathbf{E}(\tilde{\delta})$. Using Weyl’s inequality \cite{B_Bhatia97}, this eigenvalue can be lower bounded by
\begin{eqnarray}
\lambda_{+} (\tilde{\delta}) \ &\approx & \ \sigma^2_s \ \text{max} \left(  \ \text{eig} \left( \mathbf{F} - \left( 1 - \tilde{\delta} \right) \mathbf{B}  \right)  \right) \nonumber \\
&\ge & \ \Big\lfloor  \sigma_s^2 \left( \lambda_{\text{max}}^{\mathbf{F}} - \left( 1 - \tilde{\delta} \right) \ \lambda_{\text{max}}^{\mathbf{B}} \right) \Big\rfloor,
\end{eqnarray}
\noindent where $\lambda_{\text{max}}^{\mathbf{F}}$ and $\lambda_{\text{max}}^{\mathbf{B}}$ are the largest eigenvalues of the matrices $\mathbf{F}$ and $\mathbf{B}$, respectively. The outage probability depends on the eigenvalues of $\mathbf{E}  (\tilde{\delta})$ (exact expression) or the eigenvalues of $\mathbf{F}$ and $\mathbf{B}$ (lower bound). 

Assume equal power allocation, one has
\begin{eqnarray}
\lambda_{+} (\tilde{\delta}) &\approx & \frac{P_{\rm tot} \sigma_s^2}{N \left(  \sigma_s^2 + \sigma_n^2 \right)} \ \text{max} \left(  \text{eig} \left( \mathbf{r} \mathbf{r}^{H} - \left( 1 - \tilde{\delta} \right) \ \mathbf{C}  \right)   \right) \nonumber\\
&=&  \frac{ P_{\rm tot} \sigma_s^2}{ \left(  \sigma_s^2 + \sigma_n^2 \right)}  \tilde{\lambda}_{+}(\tilde{\delta})
\label{LambdaPlus} 
\end{eqnarray}
\noindent where $\tilde{\lambda}_{+} (\tilde{\delta}) =  \text{max} \left(  \text{eig} \left( \mathbf{r} \mathbf{r}^{H} - \left( 1 - \tilde{\delta} \right) \ \mathbf{C}  \right)   \right) / N$ is the factor reflecting the effect of the correlation on the largest eigenvalue and it depends on the geometry of the WSN. Considering that the term ${ P_{\rm tot} \sigma_s^2} / { \left(  \sigma_s^2 + \sigma_v^2 \right)}$ is constant for specific network setting, the factor $\tilde{\lambda}_{+}(\tilde{\delta})$ will be referred to as the normalized eigenvalue. 
The outage probability in this case is expressed as
\begin{eqnarray}
&P&_{\rm out} (\tilde{\delta})  =  1 - \frac{ \tilde{\lambda}_{+}^{N-1}}{ \prod\limits_{i, l \ne i}  ( \tilde{\lambda}_{+} - \tilde{\lambda}_i)} \nonumber\\ 
\!\!\!\!\!\!\!\! &\times &   {\rm exp} \left\{ -\frac{ \left(  \sigma_s^2 + \sigma_n^2 \right) \left( 1 - \tilde{\delta}  \right) \sigma^2_\nu}{P_{\rm tot} \sigma_s^2 \tilde{\lambda}_{+} } \right\}, 0 \le \tilde{\delta} \le 1.
\end{eqnarray}
Clearly, the outage probability is a monotonically decreasing function of $\tilde{\lambda}_{+}(\tilde{\delta})$ and thus related to the signal and observations correlation.

\section{Numerical Results}

In this section, the analytic results are confirmed by numerical simulations. Consider a WSN that consists of $N$ = 10 sensor nodes randomly located in a square area with side length of $20$~m. It is required to estimate a source parameter located $30$~m away from the center of the sensor nodes. The settings of Fig.~2 are assumed for all the following simulations unless otherwise stated. Moreover, the channel fading has variance $\sigma_g^2  = 1$. All simulation results are averaged over 1000 independent runs.

Fig.~3 shows the outage probability vs. the normalized distortion. The analytic expressions are plotted with solid, dashed, and dotted lines and the simulations are plotted with circle, diamond, and plus marks. The closed-form expression is shown to be in perfect match with the simulation results and the accuracy of the closed-form approximation of the outage probability is verified. The correlation clearly affects the outage performance where more outage occurs for both distance-based correlation models. Note that for the full-rank model, $P_{\rm out} = 1$ for $\tilde{D}$ with values less than $\approx$ 0.182 (i.e. the distortion is always larger than this value) which agrees with the results of Fig.~2.

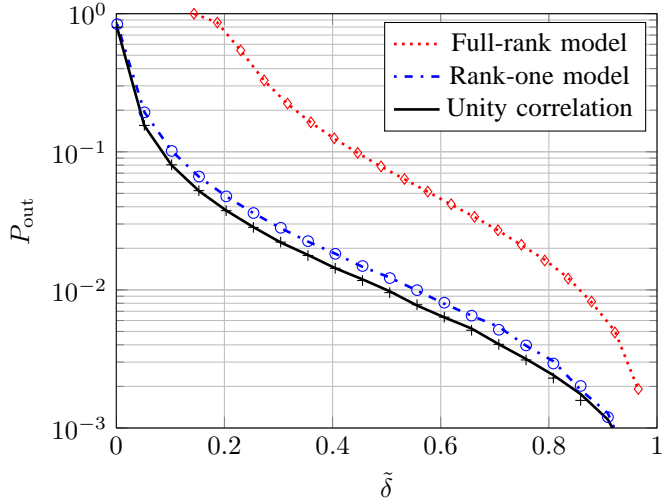
\begin{figure}[h!]
\centering
\setlength\figureheight{0.309\textwidth}
\setlength\figurewidth{0.4\textwidth}
%
%
%
%
\begin{tikzpicture}

\begin{semilogyaxis}[%
view={0}{90},
width=\figurewidth,
height=\figureheight,
scale only axis,
xmin=0, xmax=1,
xlabel={$\tilde{\delta}$},
xmajorgrids,
ymin=0.001, ymax=1,
yminorticks=true,
ylabel={$P_{\rm out}$},
ymajorgrids,
yminorgrids,
legend style={align=left}]
\addplot [
color=red,
dotted,
line width=1.0pt
]
coordinates{
 (0.143679275798458,0.998175572519084)(0.186926652969943,0.861981025371612)(0.230174030141429,0.540861187479721)(0.273421407312914,0.328049806768516)(0.3166687844844,0.222360556563816)(0.359916161655885,0.162788702026736)(0.403163538827371,0.124648810487134)(0.446410915998856,0.0980816787003629)(0.489658293170341,0.0781330208028134)(0.532905670341827,0.0633370018975329)(0.576153047513312,0.0512710144927547)(0.619400424684798,0.0415020820939904)(0.662647801856283,0.0336671368124104)(0.705895179027769,0.0269807692307673)(0.749142556199254,0.0210949248120317)(0.79238993337074,0.0165277475516862)(0.835637310542225,0.0122159383033384)(0.878884687713711,0.00814125874125493)(0.922132064885196,0.00502866242037969)(0.965379442056681,0.00189518900343266) 
};
\addlegendentry{Full-rank model};

\addplot [
color=blue,
dash pattern=on 1pt off 3pt on 3pt off 3pt,
line width=1.0pt
]
coordinates{
 (0.00177554506289581,0.83902764527892)(0.0521864649672181,0.193428937337997)(0.10259738487154,0.101426576139657)(0.153008304775863,0.0659165636588391)(0.203419224680185,0.047572484166078)(0.253830144584507,0.0359281942977826)(0.304241064488829,0.0280227987421348)(0.354651984393152,0.0223556975505852)(0.405062904297474,0.0182862595419826)(0.455473824201796,0.014694214876031)(0.505884744106118,0.0122817204301066)(0.556295664010441,0.00989510489510392)(0.606706583914763,0.00793610223642038)(0.657117503819085,0.00650357142857028)(0.707528423723407,0.00540425531914723)(0.75793934362773,0.00393805309734341)(0.808350263532052,0.00304597701149323)(0.858761183436374,0.00188950276243105)(0.909172103340696,0.00127857142857224)(0.959583023245019,0.000444444444444914) 
};
\addlegendentry{Rank-one model};

\addplot [
color=black,
solid,
line width=1.0pt
]
coordinates{
 (0.00149146491845802,0.796014322023937)(0.0519169801840858,0.154588630604817)(0.102342495449714,0.0798957123440115)(0.152768010715341,0.0521259943546268)(0.203193525980969,0.0376057649667424)(0.253619041246597,0.0284759393539843)(0.304044556512225,0.0220579576816912)(0.354470071777852,0.0179035667107006)(0.40489558704348,0.0143654743390341)(0.455321102309108,0.0119219409282694)(0.505746617574736,0.0097657430730459)(0.556172132840363,0.00771299093655464)(0.606597648105991,0.00632806324110546)(0.657023163371619,0.00525106382978568)(0.707448678637247,0.00404166666666594)(0.757874193902874,0.00315873015872981)(0.808299709168502,0.00241379310344847)(0.85872522443413,0.00176119402985064)(0.909150739699758,0.00115079365079385)(0.959576254965385,0.000491071428571743) 
};
\addlegendentry{Unity correlation};

\addplot [
color=black,
only marks,
mark=+,
mark options={solid},
forget plot
]
coordinates{
 (0.00149146491845802,0.796916772999834)(0.0519169801840858,0.155199678598198)(0.102342495449714,0.0804740229580965)(0.152768010715341,0.0523116433822401)(0.203193525980969,0.0374780673742381)(0.253619041246597,0.028367306038211)(0.304044556512225,0.0221724585289312)(0.354470071777852,0.0177318215621775)(0.40489558704348,0.0143363561704381)(0.455321102309108,0.0116910769194351)(0.505746617574736,0.00955278999698742)(0.556172132840363,0.00780711953996298)(0.606597648105991,0.00635088240261124)(0.657023163371619,0.00510346295015576)(0.707448678637247,0.00403510209325194)(0.757874193902874,0.00311001221009554)(0.808299709168502,0.00229816788813263)(0.85872522443413,0.00158418979915526) 
};
\addplot [
color=blue,
only marks,
mark=o,
mark options={solid},
forget plot
]
coordinates{
 (0.00177554506289581,0.839140491678083)(0.0521864649672181,0.193124558127731)(0.10259738487154,0.101357386178521)(0.153008304775863,0.0661640730311885)(0.203419224680185,0.0475343216588991)(0.253830144584507,0.0360300989210068)(0.304241064488829,0.0281589659325465)(0.354651984393152,0.0225407981965624)(0.405062904297474,0.018246579497921)(0.455473824201796,0.0148923535457644)(0.505884744106118,0.0121758377788332)(0.556295664010441,0.00995552999175571)(0.606706583914763,0.00807938773369475)(0.657117503819085,0.00650967555774007)(0.707528423723407,0.00514426983030447)(0.75793934362773,0.003963038005186)(0.808350263532052,0.00293074112927244)(0.858761183436374,0.00201604165090647)(0.909172103340696,0.00119788023814509) 
};
\addplot [
color=red,
only marks,
mark=diamond,
mark options={solid},
forget plot
]
coordinates{
 (0.143679275798458,0.998613757126924)(0.186926652969943,0.863175415745795)(0.230174030141429,0.542208495654733)(0.273421407312914,0.329039165756993)(0.3166687844844,0.223532301180497)(0.359916161655885,0.163475336937466)(0.403163538827371,0.125208819278034)(0.446410915998856,0.098195411432688)(0.489658293170341,0.0783789716354838)(0.532905670341827,0.0635109312604082)(0.576153047513312,0.0514165334246705)(0.619400424684798,0.041733603889912)(0.662647801856283,0.0337950019904635)(0.705895179027769,0.0270284888675732)(0.749142556199254,0.0212103213031243)(0.79238993337074,0.0163232827879079)(0.835637310542225,0.0120700028174812)(0.878884687713711,0.00822949471416765)(0.922132064885196,0.00491279592926951)(0.965379442056681,0.0019145786695712) 
};
\end{semilogyaxis}
\end{tikzpicture}%
\caption{Outage probability vs. normalized distortion ($\theta_1$ = 250, $\theta_2$ = 1, $\sigma^2_s / \sigma^2_n$ = 20~dB, $\sigma^2_x / \sigma^2_\nu$  = 20~dB, $\sigma_s^2  = 1$, and $P_{\rm tot}$ = 10 dB).}  
\end{figure}

Fig.~4 compares between the exact expression for the largest eigenvalue (normalized by $\frac{P_{\rm tot} \sigma_s^2}{  \sigma_s^2 + \sigma_n^2 }$) and its approximation in Eq.~(17) for different source locations. The source distance to the center of the square area is set to 50~m, 30~m, and 0~m. Note that $\lambda_{+}$ is upper bounded by $\frac{P_{\rm tot} \sigma_s^2}{  \sigma_s^2 + \sigma_n^2 }$ since
\begin{eqnarray}
\lambda_{+} (\tilde{\delta})&\approx & \frac{P_{\rm tot} \sigma_s^2}{N \left(  \sigma_s^2 + \sigma_n^2 \right)} \nonumber \\
&\times & \left( \text{max} \left( \text{eig} \left(   \mathbf{r} \mathbf{r}^{H} \right) \right) - \left( \! 1 \! - \! \tilde{\delta} \! \right)  \text{max} \left( \text{eig} \left(  \mathbf{C}  \right) \right) \right)  \nonumber\\
&\le &   \frac{P_{\rm tot} \sigma_s^2}{N \left(  \sigma_s^2 + \sigma_n^2 \right)}   \sum\limits_{i=1}^{N} \rho_i^2 \le \frac{P_{\rm tot} \sigma_s^2}{  \sigma_s^2 + \sigma_n^2 }.
\end{eqnarray}

Recall that the outage probability is a monotonically decreasing function of $\lambda_{+}(\tilde{\delta})$ and situations with larger eigenvalue corresponds to a better outage performance (i.e. lower outage probability). From the figure, we conclude that observing farther source reduces the largest eigenvalue and thus increases the outage probability.
%


\begin{figure}[h!]
\centering
\setlength\figureheight{0.309\textwidth}
\setlength\figurewidth{0.4\textwidth}
\input{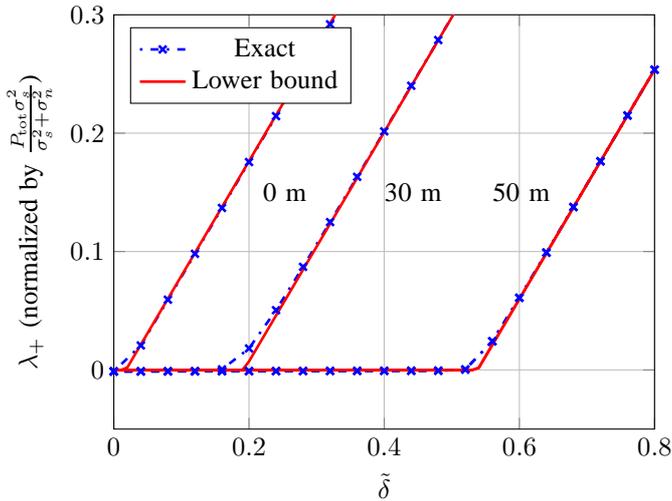}
\caption{Largest eigenvalue and its lower bound vs. normalized distortion 
($\theta_1$ = 250, $\theta_2$ = 1, $\sigma^2_s / \sigma^2_n$ = 20~dB, $\sigma^2_x / \sigma^2_\nu$  = 20~dB, $\sigma_s^2  = 1$, and $P_{\rm tot}$ = 10 dB).}  
\end{figure}

\section{Conclusions}

In this paper, distributed estimation of a scalar parameter in WSNs is considered. Correlated source signal and observations are assumed and the effect of correlation is investigated.  A closed-form expression for the outage probability is derived to link between the correlation and the outage performance. It is shown that higher distortion levels occurs with higher probability when assuming correlated observations as compared to unity correlated ones. Moreover, the distortion does not vanish when increasing the number of sensor nodes indefinitely for the distance-based correlation.

\end{document}